\documentclass[aps,prd,eqsecnum,nofootinbib,twocolumn,floatfix,superscriptaddress]{revtex4}
\usepackage{graphicx}
\usepackage{times,mathptm}
\newcommand{\beq}{\begin{equation}}
\newcommand{\eeq}{\end{equation}}
\newcommand{\bea}{\begin{eqnarray}}
\newcommand{\eea}{\end{eqnarray}}

\renewcommand{\b}{\beta}
\renewcommand{\a}{\alpha}

\newcommand{\g}{\gamma}
\newcommand{\n}{\nu}
\newcommand{\m}{\mu}

\newcommand{\bx}{{\mathbf{x}}}

\newcommand{\s}{\sigma}

\newcommand{\oh}{{\textstyle{\frac{1}{2}}}}

\newcommand{\dg}{\dagger}

\newcommand{\rf}[1]{(\ref{#1})}
\newcommand{\ra}{\rightarrow}

\newcommand{\FIGURE}[2][v]{\begin{figure}[#1]#2
              \end{figure}}

\begin{document}

\title{Vortices, Symmetry Breaking, and Temporary Confinement in SU(2)
Gauge-Higgs Theory}

\author{J. Greensite}
\affiliation{Physics and Astronomy Dept., San Francisco State
University, San Francisco, CA~94117, USA}
\author{{\v S}. Olejn\'{\i}k}
\affiliation{Institute of Physics, Slovak Academy
of Sciences, SK--845 11 Bratislava, Slovakia}
\date{\today}
\begin{abstract}
  We further investigate center vortex percolation and Coulomb gauge remnant symmetry breaking
in the SU(2) gauge-Higgs model.  We show that string breaking is visible in
Polyakov line correlators on the center projected lattice, that our usual numerical tests 
successfully relate P-vortices to center vortices, and that vortex removal 
removes the linear potential, as in the pure gauge theory.  This data suggests that global
center symmetry is not essential to the
vortex confinement mechanism.   But we also find that the line of vortex percolation-depercolation
transitions, and the line of remnant symmetry breaking transitions,
do not coincide in the SU(2)-Higgs phase diagram.  This non-uniqueness of transition lines associated with
non-local order parameters
favors a straightforward interpretation of the Fradkin-Shenker
theorem, namely: there is no unambiguous distinction, in the SU(2) gauge-Higgs models,
between a ``confining" phase and a Higgs phase.  
\end{abstract}

\maketitle
%
% Section <I
%
\section{Introduction}\label{Intro}

   Investigations of the confining force generally concentrate on gauge theories in
which confinement is permanent; i.e.\ the linear potential increases without limit.
Theories of this kind (with finite rank gauge groups) are all 
invariant under a global center symmetry, which can be expressed as
\beq
      U_0(\bx,t) \ra z U_0(\bx,t)~~~\mbox{all $\bx$, fixed $t$}
\label{ztrans}
\eeq
in lattice formulation, where $z\ne 1$ is an element of the (non-trivial) center of the 
gauge group. The unbroken realization of this
symmetry is responsible 
for the vanishing of Polyakov line expectation values, and hence permanent confinement.
For SU(N) gauge theories with this global symmetry, the
potential between static color sources, in color 
group representation $r$, depends only on the N-ality of representation $r$.  
While this fact is easily understood
in terms of energetics/string-breaking arguments (e.g.\ a flux tube between adjoint sources
can "snap" due to pair production of gluons), it also means that
the string tension of a Wilson loop, evaluated in an ensemble of configurations generated
from the pure Yang-Mills action (and therefore blind to the location of
the Wilson loop), depends only on the N-ality of the loop representation.  This leads to
a rather profound conclusion: 
large-scale vacuum fluctuations $-$ occuring in the \emph{absence} of any external
source $-$ must somehow contrive to disorder only the center degrees of freedom of Wilson loop 
holonomies.  The center 
vortex confinement mechanism (c.f.\ ref.\ \cite{review} for a review) is the simplest proposal
for how this type of disorder can occur.

   However, not all gauge theories of interest are invariant under a non-trivial
global center symmetry \rf{ztrans}.  Examples include real QCD, and any other SU(N) gauge
theory with matter fields in the fundamental representation of the gauge group.  Another
relevant example is G(2) pure gauge theory, whose center symmetry and first homotopy group 
are both trivial.\footnote{In $SU(N)/Z_N$ pure gauge theory, which has a trivial center and zero
asymptotic string tension, vortices and vortex fluctuations are no different 
from those of $SU(N)$ gauge theory, but the relevant $Z_N$ symmetry is that of the first 
homotopy group \cite{deF-J}.}  In these theories, the asymptotic string tension is zero,
and at large scales the vacuum state is similar to the Higgs phase of gauge-Higgs theory.
Such theories are examples of, rather than exceptions to, the general statement that confinement
is dependent on the existence of a non-trivial global center symmetry.
On the other hand, real QCD and G(2) pure gauge theory, as well as gauge-Higgs theory in some
regions of the phase diagram, have a static quark potential which rises linearly for some interval
of color source separation, and then becomes flat.  We will refer to this situation as 
\emph{``temporary confinement"}, reserving the term \emph{``permanent confinement"} for 
theories which have a non-zero asymptotic string tension for color sources in the fundamental 
representation.\footnote{In both cases, of course, the asymptotic particle states are color singlets.  
But this is also true in a Higgs phase, where the condensate screens any external charge.  
A similar effect occurs in electrodynamics, for electrically 
charged particles placed in a plasma or a superconductor \cite{review}.  One does not normally refer 
to electric plasmas and superconductors as confining systems; what is going on is charge screening.  
We believe it is useful to distinguish between 
this kind of screening of particle charge,
and whatever physics lies behind flux tube formation and the linear static quark potential.}

   In a theory with temporary confinement, the simple (and essentially kinematical) motivation for
the center vortex mechanism is lost.  Then it is not obvious that the center vortex picture, which
is motivated by the N-ality properties of the asymptotic string tension, is relevant.
The relevance (or irrelevance) of vortices to temporary confinement is a dynamical issue, 
which we would like to investigate via numerical simulation.  

  The simplest case to consider is SU(2) gauge-Higgs theory, with the
scalar field in the fundamental ($j=1/2$) representation. 
Two of our previous articles dealt with this model.  The first, written in collaboration with
R.\ Bertle and M.\ Faber \cite{Roman}, showed that P-vortices percolate when the couplings lie in the 
temporary confinement region of the phase diagram, and cease to percolate in the Higgs region, where there is
no linear potential at all.  We also found that center-projected Polyakov lines, in the temporary
confinement region, show evidence of color screening by the scalar field.  This work did not, however,
attempt to show that P-vortices in center-projected configurations actually correspond to center vortices in
unprojected configurations, as none of our usual tests for that correspondence were employed.  A second article,
in collaboration with D.\ Zwanziger \cite{GOZ}, considered the spontaneous breaking of a remnant global symmetry, 
which exists after Coulomb gauge fixing, in the SU(2) gauge-Higgs theory.  A confining color Coulomb potential
is associated with the unbroken realization of this remnant symmetry, and it was found that remnant
symmetry was unbroken in the temporary confinement region, and spontaneously broken in the Higgs region.  We
did not check, however, whether remnant symmetry breaking and vortex depercolation occur at
the same place in the phase diagram (although we assumed this to be true).  The reason was that the position of the
depercolation transition, found in ref.\ \cite{Roman}, was determined for a gauge-Higgs theory with variable
Higgs modulus, while the position of the remnant symmetry breaking transition, found in ref.\ \cite{GOZ}, was
computed in the frozen modulus version of the theory.  The present article is intended to fill in these
gaps in our two previous articles.

\section{Center Dominance}\label{CD}

   We consider a gauge-Higgs theory with a frozen modulus Higgs field.  For the SU(2) gauge group,
the action can be written as
\beq
    S = \b \sum_{plaq} \oh \mbox{Tr}[UUU^{\dg}U^{\dg}] + \gamma \sum_{x,\m} \oh
              \mbox{Tr}[\phi^\dg(x) U_\m(x) \phi(x+\widehat{\m})]
\label{action}
\eeq
where $\phi$ is SU(2) group-valued.  This theory was first studied numerically
by Lang et al.\ \cite{Lang}; the phase diagram is sketched in Fig.\ \ref{phase}.
There is a line of first order transitions, but only one thermodynamic phase;
any two points in the diagram can be connected by a path which avoids all non-analyticity in
the free energy.  The absence of a transition completely separating the diagram into
a confinement phase and a Higgs phase was demonstrated analytically by Fradkin and Shenker,
and Osterweiler and Seiler, in refs.\ \cite{FS}.
Nevertheless, below the transition line lies a temporary confinement
region, where the static quark potential rises linearly up to some screening distance,
while above the line the theory is Higgs-like at all distances, and the static potential
is nowhere linear.  

\FIGURE[t]{
\centerline{{\includegraphics[width=8truecm]{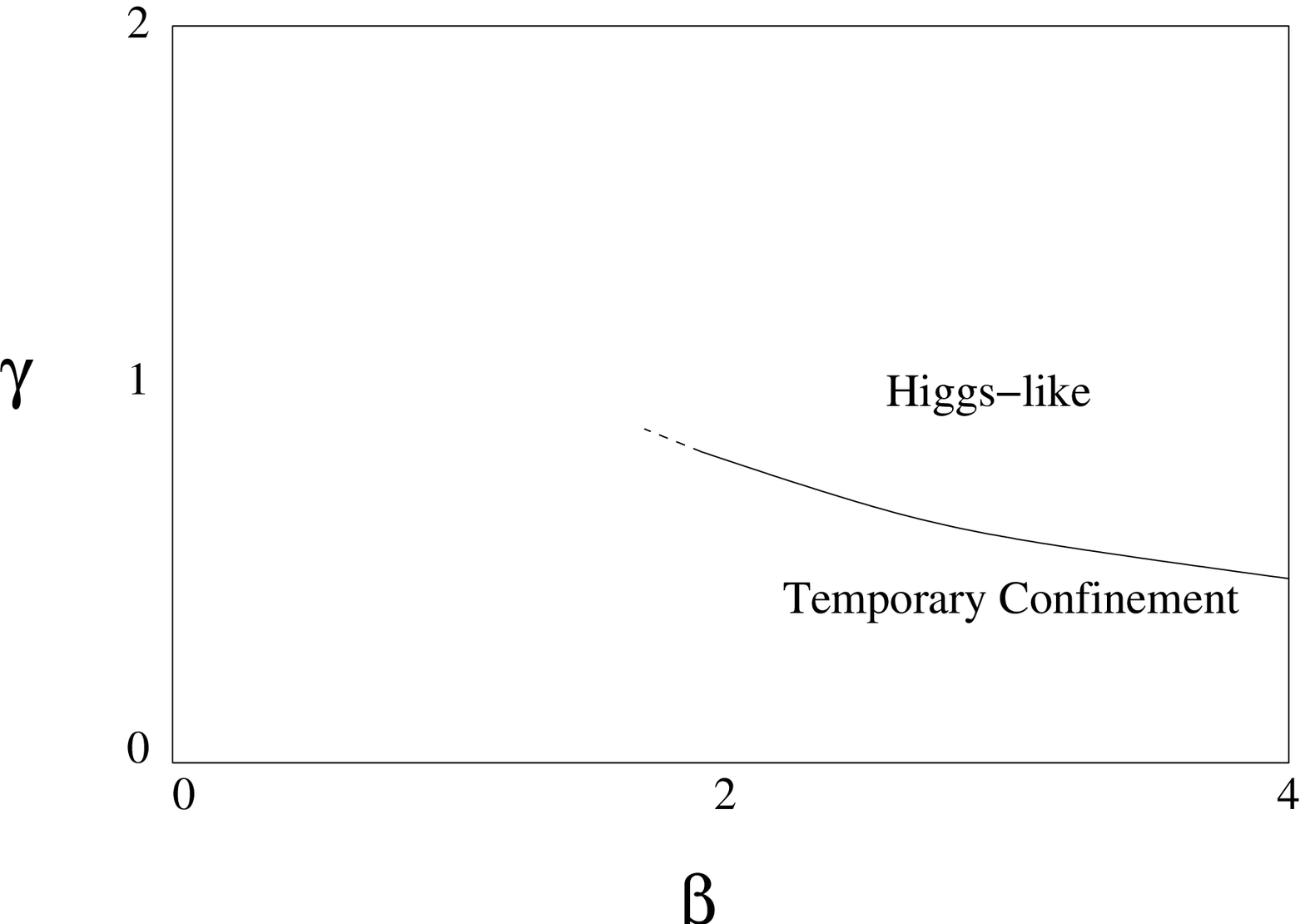}}}
\caption{Schematic phase diagram of the SU(2) gauge-Higgs system.  
The solid line is a line of first-order phase transitions.  
%The dashed line represents
%non-thermodynamic transitions in some non-local order parameter; the precise position of this line
%would depend on the order parameter chosen.
} 
\label{phase}
}

    We would like to study center dominance inside the temporary confinement region, but
close enough to the transition line so that the
the screening effect of the scalar field is detectable numerically.  For this purpose,
we compute the expectation value of Polyakov lines at $\b=2.2$, on an $L^3 \times 4$ lattice.
At $\b=2.2$, the first order transition occurs at about $\g=0.84$.
The quantity we measure is
\beq
       \langle P \rangle \equiv \langle {1\over L^3}\left|
            \sum_{\bf{x}} P({\bf x})
               \right| \rangle
\eeq
where $P({\bf x})$ denotes the Polyakov line passing through the point $\{{\bf x},t=0\}$.
In the case of unbroken center symmetry, at $\gamma=0$ and on an $L^3\times L_T$ lattice, 
we must find 
\beq
     \langle P \rangle  \propto \sqrt{1\over L^{3}}
\label{P}
\eeq
while for explicitly broken center symmetry ($\g\ne 0$) it must be that $\langle P \rangle$ has
a non-zero limit at large volume.

  Our data for Polyakov lines on the unprojected lattice, at $\b=2.2$ and $\g=0,~0.71$, is shown if Fig.\
\ref{p00}, for lattice sizes up to $20^3 \times 4$.  The straight line is a best fit through the
$\g=0$ data, and errorbars for some data points are smaller than the
symbol size.  It is clear that the $\g=0$ data is consistent with
eq.\ \rf{P}, and $\langle P \rangle$ extrapolates to zero in the
infinite volume limit.  At $\g=0.71$ the system is still below the first-order transition line, and
in the temporary confinement region.  It appears from the data that at this coupling,  $\langle P \rangle$ 
has stabilized (at $L=14,16,20$) to a non-zero value of $\langle P \rangle \approx 0.034(1)$.  
So at $\b=2.2,~\g=0.71$, color screening of Polyakov lines by the matter field is detectable.  

   The data at these same couplings, for Polyakov lines on the center projected lattice is displayed
in Fig.\ \rf{zp00}.  The center projected data tells exactly the same story as the data on the unprojected
lattice: at $\g=0$, Polyakov lines tend to zero at large volumes, while at $\g=0.71$ screening is detected,
and the Polyakov lines stabilize at $\langle P \rangle \approx 0.120(4)$.  This aspect of center dominance
in the SU(2) gauge-Higgs model was previously found in ref.\ \cite{Roman}, for the variable modulus version
of the theory.

\FIGURE[tbh]{
\centerline{{\includegraphics[width=8truecm]{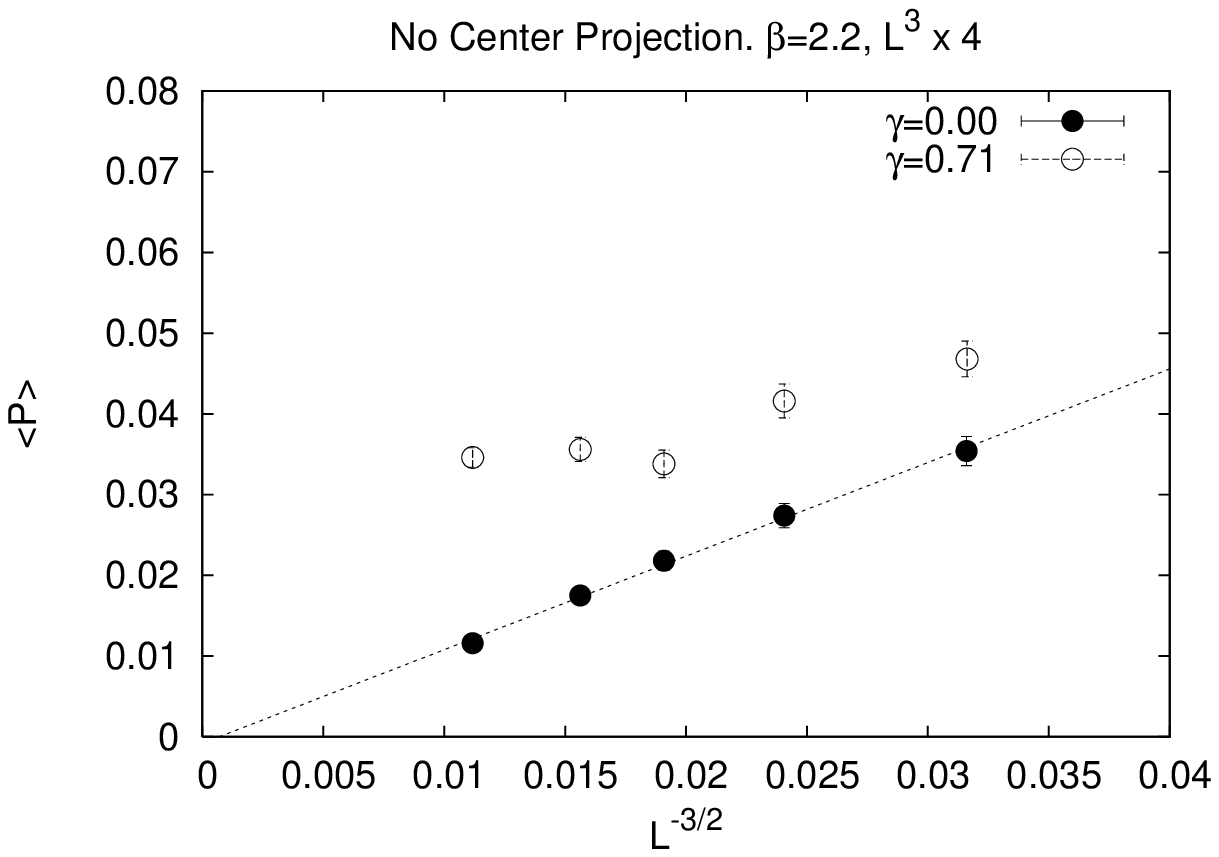}}}
\caption{Polyakov line values, on the unprojected lattice at $\b=2.2$,
$\gamma=0$ and $\g=0.71$, on $L^3\times 4$ lattice volumes with
$L=10,12,14,16,20$  The straight line is a best fit to the $\g=0$ data.} 
\label{p00}
}

\FIGURE[tbh]{
\centerline{{\includegraphics[width=8truecm]{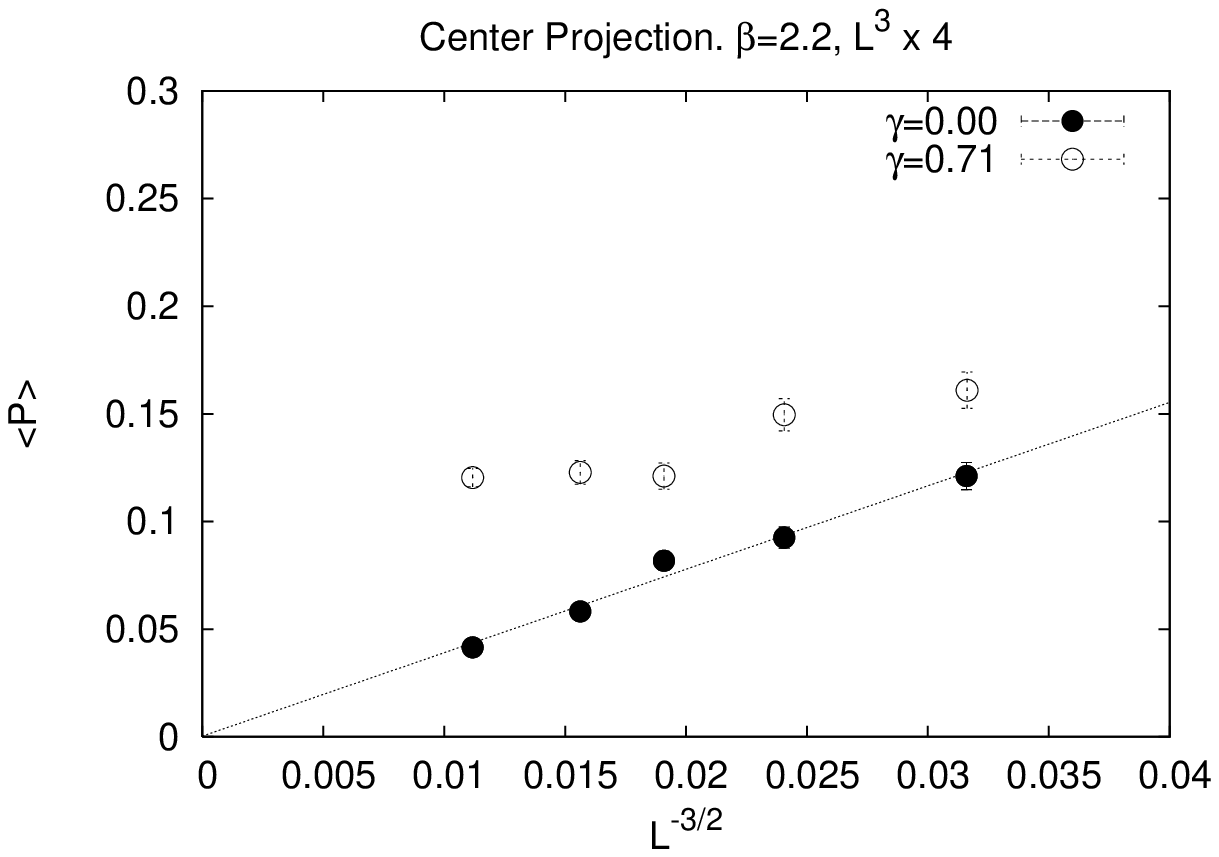}}}
\caption{Same as Fig.\ \ref{p00}, for Polyakov lines on the
center-projected lattice.} 
\label{zp00}
}

  We can go on to calculate the correlator of center-projected Polyakov lines
$\langle P(x) P(x+R) \rangle$
at $\b=2.2,~\gamma=0.71$, on a $20^3 \times 4$ lattice.  The data is
shown in Fig.\ \ref{zpoly}.  The dashed line is a best fit to the data,
for $R \ge 2$, by the function
\beq
       f(R) = c_0 + c_1 \exp[-4\s R]
\eeq
From the fit we find $c_0=0.0182 ,~\s=0.211$.  Not surprisingly,
$c_0$ is quite close to the square of the VEV of the
Polyakov line in center projection, which is $\langle P_{cp} \rangle = 0.12$
on the $20^{3}\times 4$ lattice.   In this way we
see string-breaking, due to the dynamical matter field, from Polyakov
line data on the center-projected lattice. 

\FIGURE[tbh]{
\centerline{{\includegraphics[width=8truecm]{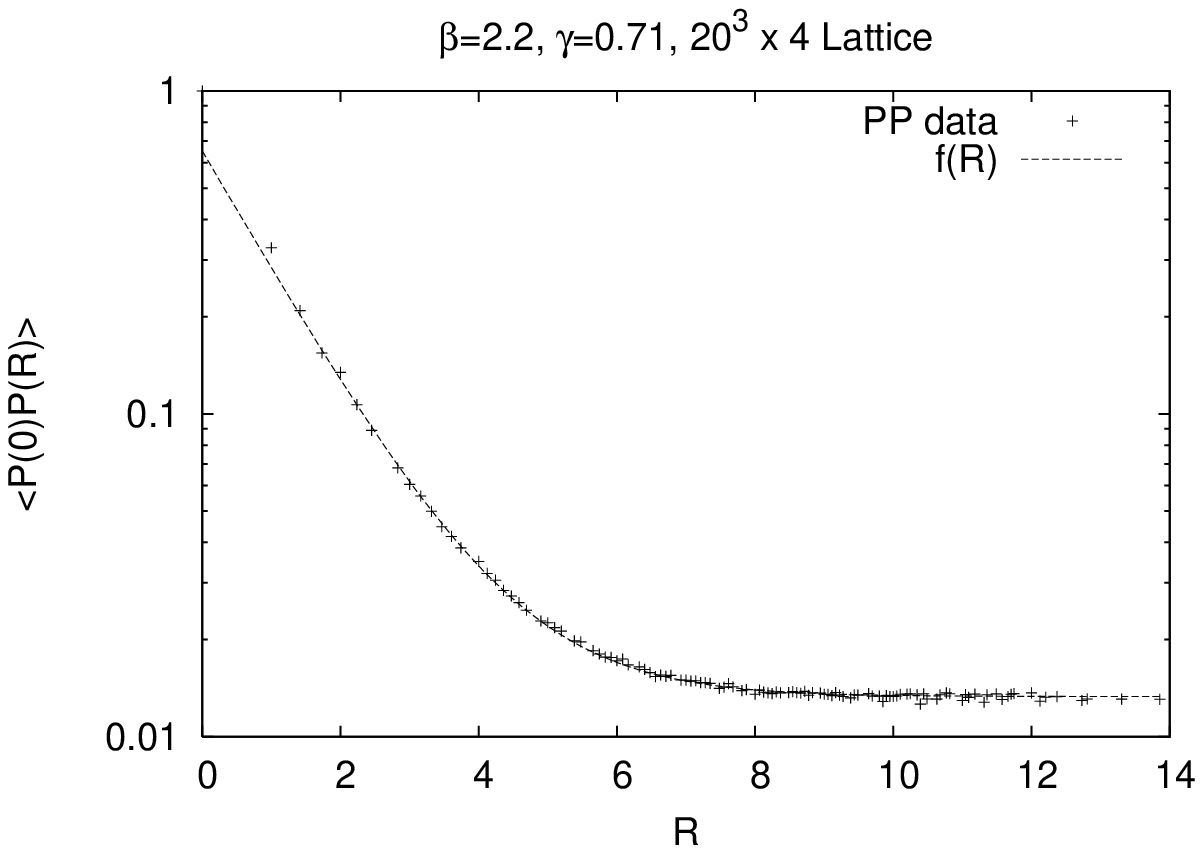}}}
\caption{Polyakov line correlator $\langle P(0)P(R)\rangle$ on the center-projected lattice.} 
\label{zpoly}
}
 
   Although the data displayed in the previous graphs makes a good case for 
center dominance in Polyakov lines in gauge-Higgs theory (which is not a new result), 
there is still the question of whether P-vortex excitations
on the center-projected lattice correlate with gauge-invariant observables on the
unprojected lattice.  At $\g>0$ global center symmetry is broken, and the 't Hooft loop operator
$B(C)$ \cite{thooft} which creates a thin center vortex would not only raise the action 
at the loop location, but also on some surface bounded by the vortex loop.  The position
of this "Dirac surface" is no longer a gauge artifact.   We can still identify P-vortices
via maximal center gauge fixing and projection, but the correspondence of these P-vortices
to center vortices on the unprojected lattice cannot be taken for granted.   
Our standard test for this correspondence is to see if $W_1(C)/W_0(C) \ra -1$ in the
large-loop limit. Here $W_n(C)$ represents a Wilson loop, computed from unprojected link
variables, with the restriction that the minimal area of loop $C$ is pierced by $n$ P-vortices
on the projected lattice.  The result of this test, for spacelike loops on a $20^3\times 4$
lattice at $\b=2.2,~\g=0.71$ is shown in Fig.\ \ref{w01}.  It is much
like the result found for pure gauge theories, and seems
perfectly consistent with the assumed correspondence of P-vortices and center vortices.
 
\FIGURE[tbh]{
\centerline{{\includegraphics[width=8truecm]{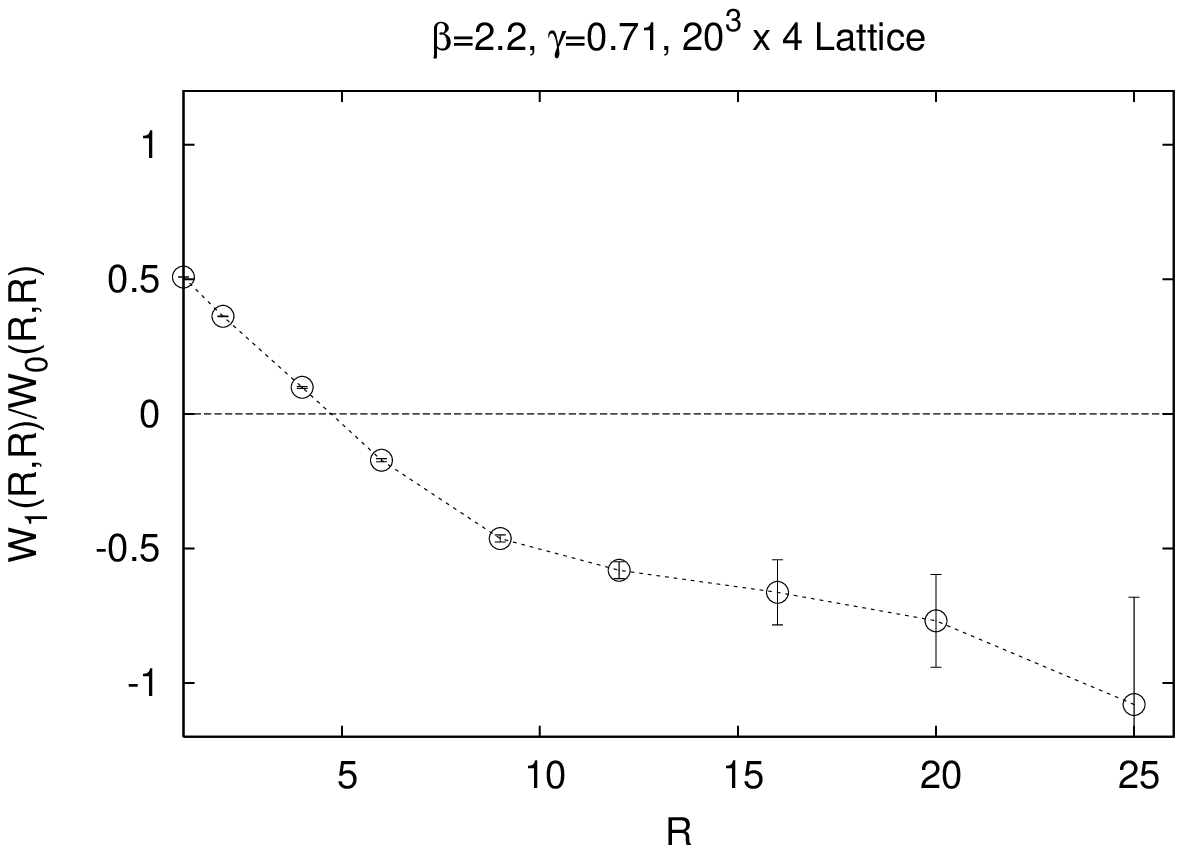}}}
\caption{Ratio of "vortex-limited" Wilson loops.  $W_1(C)$ is evaluated for loops pierced
by a single P-vortex; $W_0(C)$ is evaluated for loops which are not pierced by any P-vortices.} 
\label{w01}
}

\FIGURE[tbh]{
\centerline{{\includegraphics[width=8truecm]{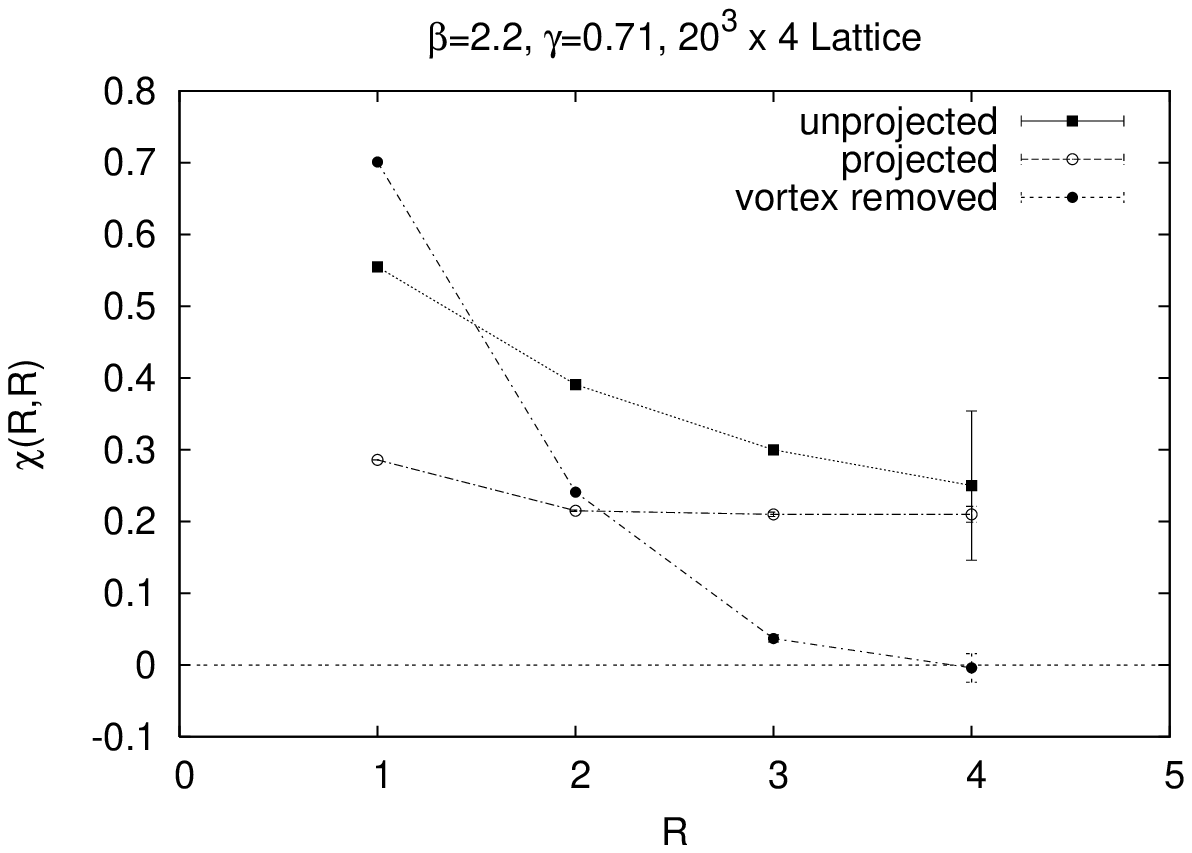}}}
\caption{Creutz ratios in the gauge-Higgs theory for unprojected, center-projected,
and vortex-removed lattices.} 
\label{chi}
}

   Finally, we compare the Creutz ratios of projected and unprojected spacelike
Wilson loops at $\b=2.2,~\g=0.71$, on the $20^3 \times 4$ lattice, and look for the effect of
vortex removal.  The relevant data is displayed in Fig.\ \ref{chi}.  We see that the projected
Creutz ratios are constant for any $R>1$, as in the pure-gauge theory, at roughly 
$\chi(R,R) \approx 0.21$, and the values for the Creutz ratios on the unprojected lattice
also appear to converge towards this value.  Note that this value for the asymptotic string
tension is consistent with the value obtained from Polyakov line correlators on the projected
lattice.  The effect of vortex removal is also shown in Fig.\ \ref{chi}.  Vortices are removed
via the de Forcrand-D'Elia prescription \cite{dFE}, which consists of fixing to maximal center gauge, 
and multiplying each link variable by its center-projected value.  This is the vortex-removed ensemble.
We see that the Creutz ratios in this ensemble go to zero asymptotically, just as in the
pure gauge theory.

\section{Symmetry Breaking and Vortex Percolation}\label{Perc}

    The Fradkin-Shenker theorem \cite{FS} assures us that there is no phase transition which
completely isolates the temporary confinement region from a Higgs phase; at least,
no such transition could be detected by any local order parameter.  But what about
non-local order parameters?  Perhaps thermodynamics is not the ultimate arbiter, and
there really exists some qualitative difference between the temporary confinement and
Higgs phases, characterized by symmetry-breaking, or by condensation of solitonic objects,
which is only detectable via non-local observables.  A relevant example is the Ising model in the
presence of a small external magnetic field $h$.  In that case the global $Z_{2}$ symmetry
of the zero-field model is explicitly broken, and there is no thermodynamic transition
from an ordered to a disordered state.  On the other hand, there is a sharp depercolation
transition in the $h>0$ as well as the $h=0$ case; the line of such transitions
in the temperature-$h$ phase diagram is known as a {\sl Kert{\'e}sz line} \cite{Kertesz}.
In the gauge-Higgs model the coupling $\g>0$ breaks the global $Z_2$ symmetry, and it is
possible that a sharp vortex depercolation transition could
serve to distinguish the temporary confinement and Higgs phases of the theory.

    An alternative proposal for distinguishing these phases is associated with symmetry
breaking. We know from the Elitzur theorem that a local gauge symmetry can never be
spontaneously broken.  On the other hand, after Coulomb or Landau gauge fixing there still exists some
global remnant of the local symmetry, and these global symmetries \emph{can} be
spontaneously broken.  So perhaps center vortex depercolation and remnant symmetry
breaking define a unique Kert{\'e}sz line, which unambiguously separates the temporary 
confinement and Higgs phases of the gauge-Higgs model \cite{Kurt}.  
Another candidate symmetry for distinguishing the two phases, advocated by the Pisa group, is
a certain dual (abelian) magnetic symmetry \cite{Adriano}.  This approach, in the non-abelian
theory, also requires fixing to some gauge.

    Since both the identification of vortices, and the definition of remnant (as well as magnetic)
symmetries entails the choice of a gauge,  the associated order parameters are non-local (if 
expressed as gauge-invariant operators),
and the Fradkin-Shenker theorem does not rule out non-analytic behavior in such observables.  
On the other hand, if the transition lines associated with each order parameter do not coincide, 
then the claim that
any of these parameters can be used to ``define" confinement, in the absence of a non-vanishing
asymptotic string tension, becomes less compelling.    In this section we will report on our
results for the Kert{\'e}sz lines corresponding to vortex depercolation, and to remnant symmetry
breaking in Coulomb gauge.

\subsection{Symmetry Breaking}
    We begin by reviewing some points made in ref.\ \cite{GOZ}.  First of all, there is
a remnant symmetry in "minimal" Coulomb gauge, defined as the
gauge with minimizes, on the lattice,
\beq
            R = - \sum_{x} \sum_{k=1}^{3} \mbox{ReTr}[U_{k}(x)]  
\eeq
Fixing to this gauge still allows the following "remnant" gauge transformations which are global 
in space, but local in time:
\beq
           U_{k}(\bx,t) = g(t)U_{k}(\bx,t)g^{\dg}(t) ~~~,~~~ 
           U_{0}(\bx,t)  = g(t)U_{0}(\bx,t)g^{\dg}(t+1)
\label{remnant}
\eeq
On any given time slice, this global symmetry can be spontaneously broken, and such breaking
implies the absence of a confining color Coulomb potential.  The color Coulomb
potential can be extracted, at weak couplings, from the correlator of timelike links at a given time, 
i.e.\ \cite{GOZ}
\beq
             V_{c}(R) = -\log\left[\left\langle \oh \mbox{Tr}[U_{0}(0,t)]U^{\dg}_{0}(R,t)]\right\rangle \right]
\eeq
$V_c(R)$ converges to the instantaneous color Coulomb potential in the continuum limit.
Asymptotically, this potential is also an upper bound on the static quark potential \cite{Dan}
\beq
              V(R) \le V_{c}(R)
\eeq
so that a confining Coulomb potential is a necessary, but not a sufficient, condition for permanent
confinement.  
    
    The remnant symmetry breaking order parameter $Q$ is expressed in terms of the timelike link variables
averaged, at a given time, over spatial volume ($L^3$)
\beq
            \widetilde{U}(t) = {1\over L^3} \sum_{\bx} U_0(\bx,t)
\eeq
If the remnant symmetry \rf{remnant} is unbroken, then the modulus of $\widetilde{U}$ should vanish in
the infinite volume limit, at any $t$.  We therefore define the order parameter as
\beq
          Q = {1\over L_t}\sum_{t=1}^{L_t} \left\langle \sqrt{\oh 
                 \mbox{Tr}[\widetilde{U}(t) \widetilde{U}^\dg(t)} \right\rangle
\eeq
where $L_t$ is the lattice extension in the time direction. $Q$ vanishes in the large volume limit in the
unbroken phase, and has a non-zero limit if the remnant symmetry is spontaneously broken.

An exponential
falloff in the timelike link correlator implies that the color Coulomb potential rises linearly
with separation.  In contrast,
if remnant symmetry is broken spontaneously, then $V_{c}(R) \ra$ constant as $R\ra \infty$.  So the
existence of an asymptotic Coulomb string tension $\s_{coul}>0$ depends on
the unbroken realization of remnant gauge symmetry, and this fact makes the $Q$ 
order parameter a good
candidate for isolating the temporary confinement phase from the Higgs phase.

\FIGURE[tbh]{
\centerline{{\includegraphics[width=8truecm]{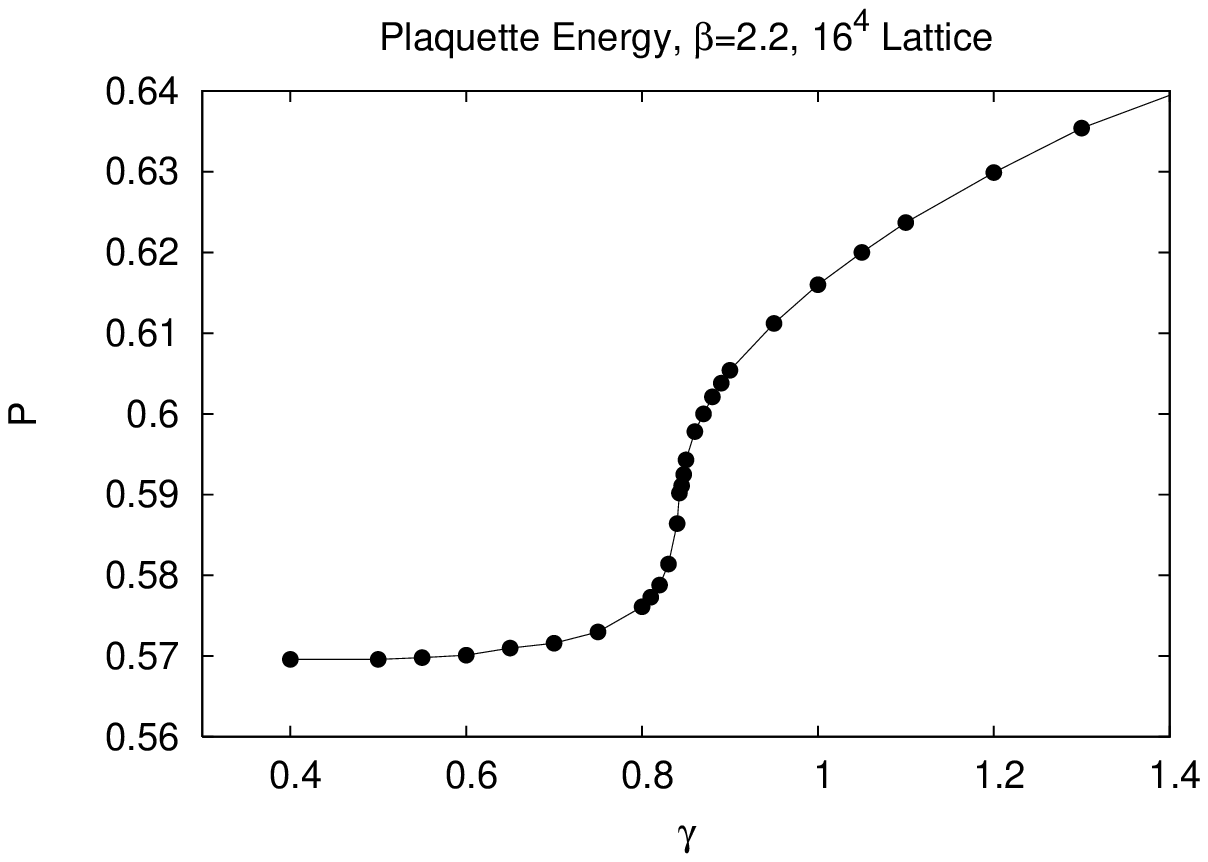}}}
\caption{Plaquette energy $P$ vs.\ Higgs coupling $\g$ at $\b=2.2$;
a weak first-order transition is seen.} 
\label{2p2e}
}

\FIGURE[tbh]{
\centerline{{\includegraphics[width=8truecm]{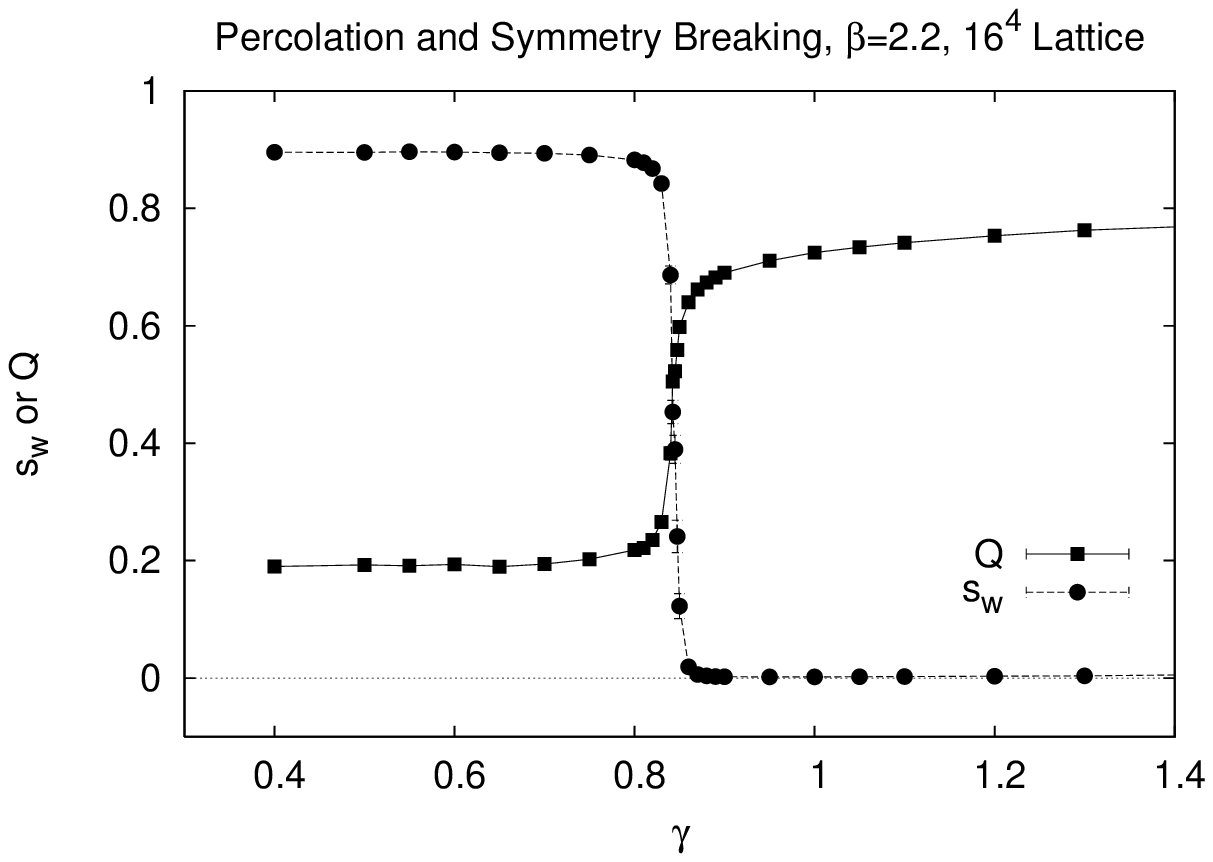}}}
\caption{Data for the remnant symmetry order parameter $Q$ vs.\ $\g$, and
percolation order parameter $s_{w}$ vs.\ $\g$, at $\b=2.2$.} 
\label{2p2}
}

       Further support for this view of $Q$ as an order parameter comes from the fact
that that the $Q$-transition line in the gauge-Higgs theory coincides with the 
thermodynamic first-order phase
transition line, up to the terminating point of that line of transitions.  
Like the plaquette energy, the data suggests that $Q$ is discontinuous along the transition line
in the infinite volume limit.  In Fig.\ \ref{2p2e} we show the
plaquette energy curve, as a function of $\g$, at $\b=2.2$;   a weak first-order transition
is visible near $\g=0.84$.\footnote{We cannot rule out the possibility that this is a very sharp
crossover, rather than an actual first-order transition.}  
Fig.\ \ref{2p2} shows our data for $Q$ vs.\ $\g$, again at $\b=2.2$,
along with another observable $s_{w}$ to be discussed shortly.  A sudden
jump in $Q$ is visible at the same value of $\g$ (within our resolution) that the jump in
plaquette energy is observed.  The non-zero value of $Q$ below the transition is a finite-size
effect; this quantity should vanish on an infinite lattice \cite{GOZ}.   

    As $\b$ is reduced, the first-order transition disappears, as seen in the plot of plaquette
energy vs.\ $\g$ at $\b=1.2$, shown in Fig.\ \ref{1p2e}.  There is still, however, a transition 
in $Q$, as seen in Fig.\ \ref{1p2}.  The $Q$ order parameter is not discontinuous in this case, instead, 
$Q$ increases continuously away from zero (in the infinite volume limit) upon
crossing the transition line, at about $\g=1.5$.  This behavior is reminiscent of magnetization in a spin system, in
the neighborhood of a second order phase transition.
In Fig.\ \ref{1p2} we show data for both $8^{4}$ and $16^{4}$ lattices,
to show the trend to $Q=0$, at infinite volume, below the transition.  We have determined the position
of the remnant symmetry-breaking transition line at a range of couplings below $\b=2.2$; this is the
lower line shown in Fig.\ \ref{gcrit}.\footnote{The Q-transition line shown in Fig.\ \ref{gcrit}
differs somewhat in location from the line we reported previously in Fig.\ 12 of ref.\ \cite{GOZ}.  The calculation
of that figure suffered from an unfortunate program error; our current Fig.\ \ref{gcrit} corrects and replaces it.  
We note that this program error did not affect the other results reported in ref.\ \cite{GOZ}.}  

\FIGURE[tbh]{
\centerline{{\includegraphics[width=8truecm]{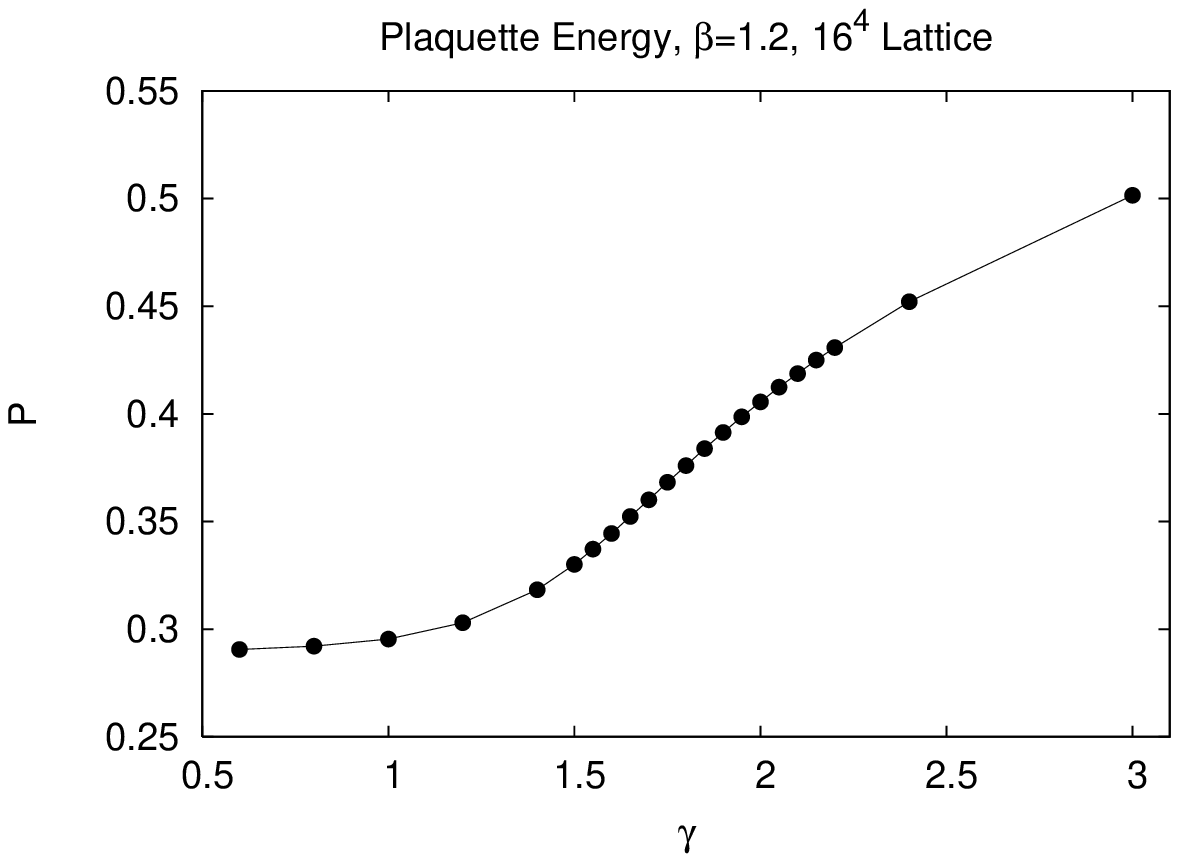}}}
\caption{Plaquette energy $P$ vs.\ Higgs coupling $\g$ at $\b=1.2$;
no transition is evident.} 
\label{1p2e}
}

\FIGURE[tbh]{
\centerline{{\includegraphics[width=8truecm]{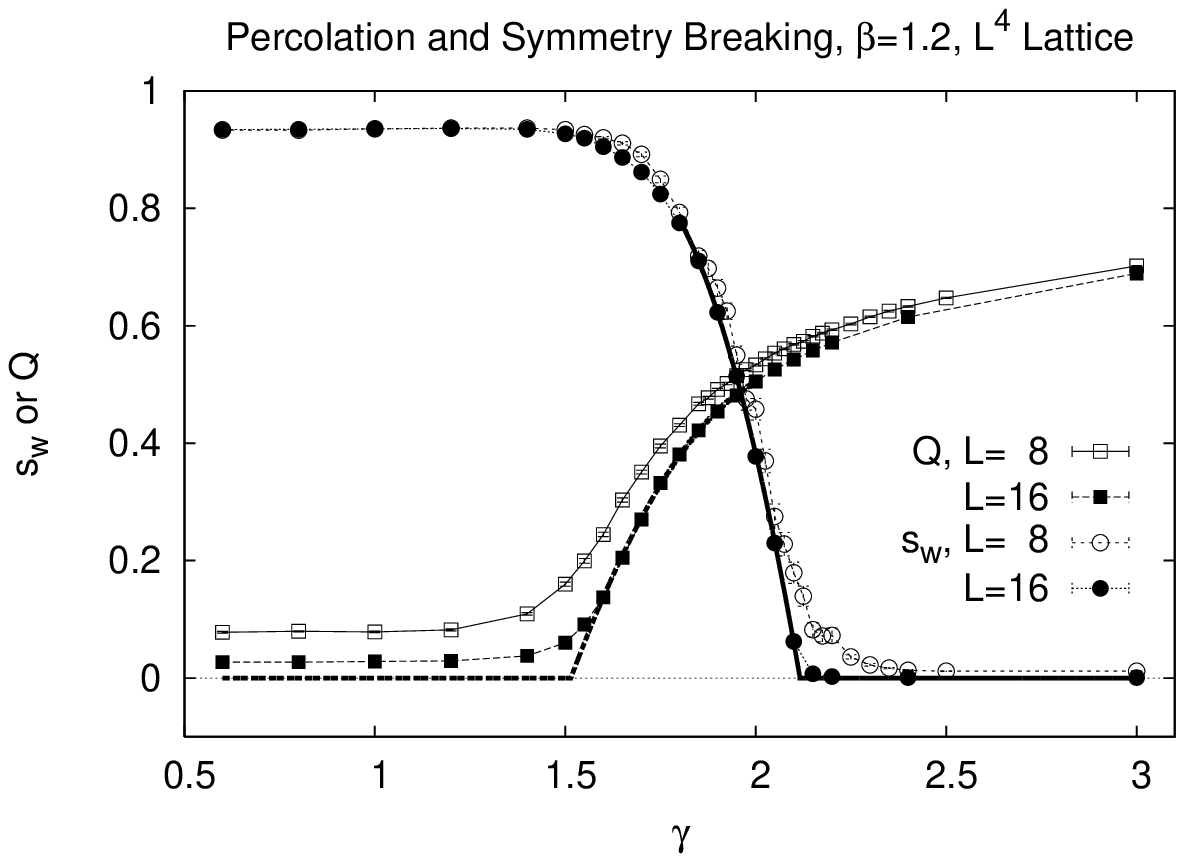}}}
\caption{At $\b=1.2$, the remnant symmetry and depercolation
transitions occur at different values of $\g$.  Solid lines are
presumed extrapolations to infinite volume.} 
\label{1p2}
}
 
\subsection{Vortex Percolation}

    We also have a second candidate for the role of order parameter, distinguishing between the
temporary confinement and Higgs phases.  This is an operator which is sensitive to the vortex
percolation-depercolation transition \cite{Roman}, denoted $s_w$, and defined as follows:  Let
$f(p)$ be the fraction of the total number $N_P$ of P-plaquettes on the lattice, carried by the P-vortex containing
the P-plaquette $p$.  Then $s_w$ is the value of $f(p)$ when averaged over all P-plaquettes.  It
can be thought of as the fraction of the total number of P-plaquettes on the lattice, contained in the ``average"
P-vortex.  More precisely: let the index $i=1,2,...,N_v$ denote vortex number, and $i(p)$ specifies
the vortex containing the P-plaquette $p$.  Also let $n_i$ denote the total number of P-vortices contained
in vortex $i$.  Then
\beq
       s_w \equiv {1\over N_P} \sum_{p=1}^{N_p} {n_{i(p)}\over N_P} = \sum_{i=1}^{N_v} {n_i^2 \over N_P^2}
\eeq
If all P-plaquettes belong to a single vortex, then $s_w=1$.  In the absence of percolation, the fraction
of the total number of P-plaquettes carried by any one vortex vanishes in the infinite volume limit.
If a finite fraction of P-plaquettes is carried by a finite number of percolating vortices in the same
limit, then $s_w>0$.  The transition from $s_w>0$ to $s_w=0$ in the large volume limit identifies the
percolation-to-depercolation transition.
In ref.\ \cite{Roman} we determined the line of depercolation transition in the gauge-Higgs model with variable Higgs
modulus; here we report the location (upper line in Fig.\ \ref{gcrit})
in the gauge-Higgs model \rf{action}, for comparison with
the remnant symmetry-breaking line (lower line in Fig.\ \ref{gcrit}).
The calculation of $s_{w}$ requires identifying, on each center-projected lattice configuration,
the number of separate P-vortices and the area of each.   Our algorithm for doing this
is described in detail in the Appendix.

\FIGURE[tbh]{
\centerline{{\includegraphics[width=8truecm]{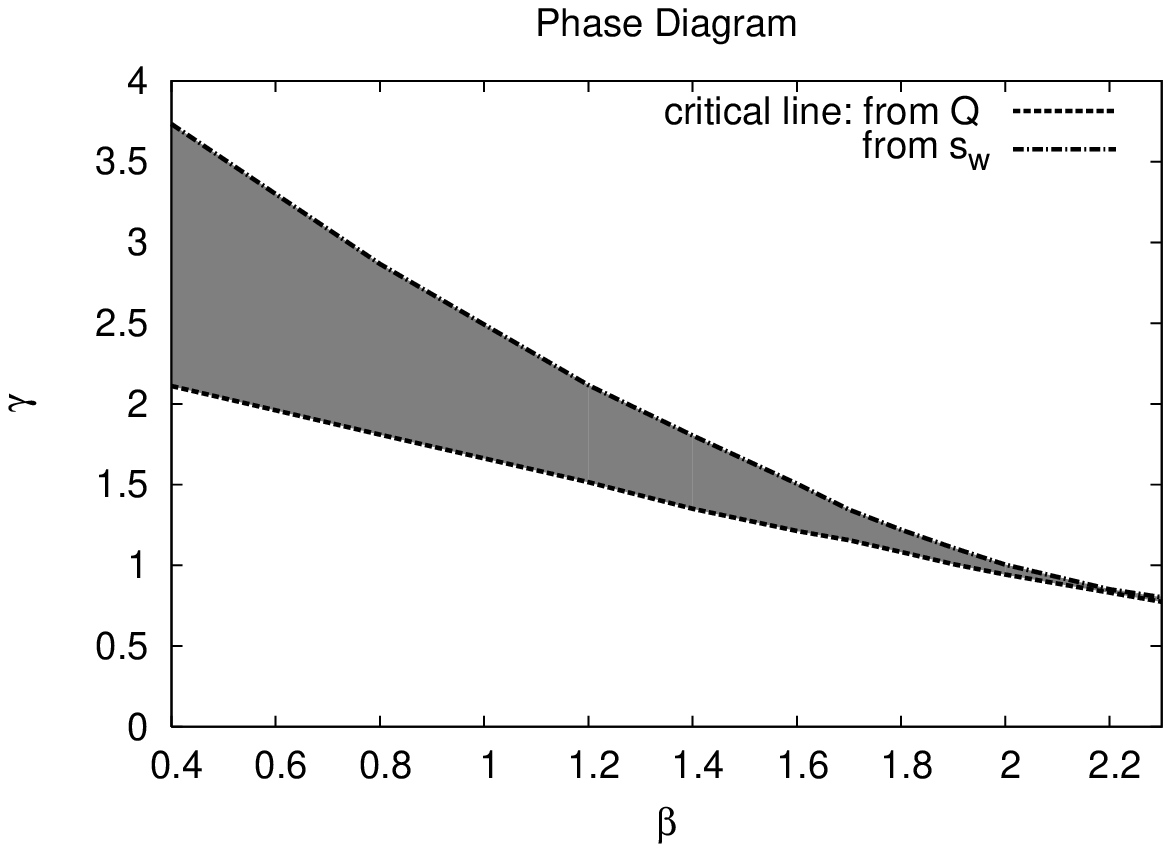}}}
\caption{Transition lines for the remnant symmetry breaking transition 
(lower line, order parameter $Q$) and the percolation-depercolation transition
(upper line, order parameter $s_{w}$).  The shaded region is a region of couplings
where remnant symmetry is broken, but vortices still percolate.  The two transitions
lines appear to converge at the end-point of the line of first-order thermodynamic
transitions.} 
\label{gcrit}
}

     The vortex depercolation transition $s_{w}\ra 0$, like the $Q$-transition,
coincides with the line of thermodynamic, first-order transitions up to the
endpoint of that line.  The data for $s_w$ vs.\ $\g$ at $\b=2.2$ is displayed, together
with the $Q$ data, in Fig.\ \ref{2p2}.  However, beyond the first-order transition
line, the vortex depercolation and remnant symmetry-breaking transitions no longer
coincide, as is evident from our data for $Q$ and $s_{w}$ at $\b=1.2$,
displayed in Fig.\ \ref{1p2}.  At this coupling, the $Q$ transition occurs at about
$\g=1.5$, while $s_{w}$ goes to zero at $\g=2.1$.

    The transition lines for both remnant-symmetry breaking
and depercolation can be compared in Fig.\ \ref{gcrit} over a range of couplings $\b$, 
and it is evident that these transition lines do not coincide,
contrary to what was assumed implicitly in ref.\ \cite{GOZ}.   There is a region between
the two transition lines where vortices percolate, but $\s_{coul}=0$.  This is 
consistent with the notion that vortex percolation is a necessary, but not
sufficient, condition for having a confining Coulomb potential, which is itself a
necessary, but not sufficient, condition for permanent confinement.

   The fact that the vortex and remnant-symmetry transition lines do not
coincide tends to support the most straightforward interpretation of the
Fradkin-Shenker theorem, namely, that there is no unambiguous distinction
between the temporary confinement and Higgs phases.  In either region, the large-scale
gauge-field fluctuations responsible for disordering Wilson loops are suppressed, and
the gauge field due to an external static source falls off exponentially with distance
from the source.  In this sense the regions are very much alike in the far infrared, 
and are characterized by charge screening, rather than confining forces.

\section{Conclusions}

     There are two conclusions.  First, the vortex mechanism for producing a
linear potential can work even when the gauge action does not possess a global
center symmetry, and the static potential is flat at large distance scales.
Thus global center symmetry is not \emph{necessarily} essential to the vortex
mechanism.  The same distribution of vortices which produces a linear potential 
over a finite interval, in temporary confinement theories, can also avoid producing 
a linear potential at asymptotic distances, as
seen in the Polyakov line correlator on the center projected lattice.  Of course, the distribution
of percolating P-vortices responsible for permanent confinement at $\g=0$, and that responsible for
temporary confinement at $\g>0$, must differ qualitatively in some way at large scales. 
In the latter case, we would expect that vortex piercings of the minimal area of a very large
Wilson loop would tend to come in pairs, whose effect on the large loop would cancel.  Whether
this effect is due to vortices having a branched polymer structure at large scales, or is due to 
some other distribution, is left for future investigation.

    The second conclusion concerns the question of whether it is possible, in a theory
without a local order parameter for confinement, to nonetheless distinguish between a
``confined" phase and a Higgs phase via some non-local order parameter.  We have 
investigated two reasonable candidates: (i) the $Q$ observable which tests for spontaneous
breaking of remnant gauge symmetry in Coulomb gauge, corresponding to the loss of
a confining color Coulomb potential, and (ii) the $s_w$ observable, which
tests for P-vortex percolation.   Both of these observables are closely related to confinement
in pure gauge theories; unbroken remnant symmetry is a necessary condition for 
permanent confinement, and vortex removal removes the confining force.  Moreover,
both observables have a transition in the gauge-Higgs phase diagram which agrees with
the first-order transition line, up to the endpoint of that line.  Beyond the first-order transition
line, however, we find that the remnant-symmetry breaking and vortex depercolation lines
do not coincide, which means that the separation of the gauge-Higgs phase diagram into a
"confinement" phase and a Higgs phase is ambiguous.   The choice of a particular non-local
observable to be an order parameter for confinement is not very compelling, if the only
non-analytic behavior seen at the transition is in that particular observable.  The fact is that
throughout the phase diagram, the gauge-Higgs model at large scales is best described
as a color-screening phase.
In this model there are no large-scale gauge field fluctuations, characteristic of confinement, 
which disorder Wilson loops, and the color field due to a static source is screened 
(as in an electrically charged plasma, or in an electric superconductor), rather than collimated into a 
flux tube.  This observation, together with our numerical result, tends to support the most 
straightforward reading of the Fradkin-Shenker
theorem: There is no essential distinction, in a gauge-Higgs model, between the temporary
confinement phase and the Higgs phase.

\acknowledgments{%
J.G.\ would like to thank Jan Ambj{\o}rn, Phillipe de Forcrand, and Jerzy Jurkiewicz,
for helpful discussions.  We would also like to thank Roman Bertle, Manfried Faber,
and Daniel Zwanziger for collaboration on articles related to the present work.
Our research is supported in part by the U.S.
Department of Energy under Grant No.\ DE-FG03-92ER40711 (J.G.) and the Slovak Science
and Technology Assistance Agency under Contract No.\  APVT-51-005704 (\v{S}.O.).
}

\appendix*
\section{}
  In this appendix we describe our procedure for identifying individual P-vortex surfaces.
The basic idea is that two P-vortex plaquettes on the dual lattice which share a common link
must belong to the same P-vortex.  An ambiguity arises when four or six plaquettes share a
link.  This could be a self-intersection of a single P-vortex, or an intersection of two or more
separate P-vortices.  We simply ignore these
ambiguous links; they are not used to identify different plaquettes as belonging to the same vortex
surface.  The algorithm goes as follows:
\begin{description}      
\item{I.}  Gauge fix the SU(2) lattice configuration to maximal center gauge, and center project.
The center-projected plaquettes have values $z_{\m\n}(x)=\pm 1$, where $x$ is the lattice site,
and  $\m,\n$ specifies the plane of the plaquette.
\item{II.} Map the above plaquette variables onto variables onto plaquette variables of the
dual lattice
\beq
         z^{D}_{\a\b}(x+\widehat{\m}+\widehat{\n}) = z_{\m\n}(x)
\eeq
Although by convention the coordinates of points on the dual lattice are half-integer, we have
added a constant vector $(\oh,\oh,\oh,\oh)$ to all dual lattice sites in order have integer
coordinates on the dual lattice also.
\item{III.}  Count the number of negative plaquettes on the dual lattice, and assign to
each of these a number $n_{\a\b}(x)$ from 1 to $A$, where $A$ is the total number of
negative plaquettes.
\item{IV.} Initialize $n_l=0$.  Loop through all of the links of the dual
lattice.  For each link shared by two \emph{and only two} negative
plaquettes, increment $n_l$, and store the plaquette numbers of the two
negative plaquettes in $p(n_l,1),p(n_l,2)$.  We will refer to such links as
"surface-pair" links.   Upon completion of the loop over links,
set $N_l$ equal to the final value of $n_l$; this is the total number of
surface-pair links. 
\item{V.} Initialize the Vortex Number of each negative plaquette,
$V(n) = 0,~n=1,...,A$, and set $n_v=0$.  Now loop through surface-pair links,
$n_l=1,...,N_l$. At each such link denote
$p_1=p(n_l,1),p_2=p(n_l,2)$, and then perform the following operation on
the vortex numbers:
\begin{enumerate}
\item if $V(p_1)=V(p_2)=0$, increment $n_v \ra n_v+1$, and set
$V(p_1)=V(p_2)=n_v$.
\item if $V(p_1) \ne 0,~V(p_2)=0$, set $V(p_2)=V(p_1)$.
\item if $V(p_2) \ne 0,~V(p_1)=0$, set $V(p_1)=V(p_2)$.
\item if both $V(p_1),~V(p_2$) are non-zero, and $V(p_1)<V(p_2)$,
set $V(p_2)=V(p_1)$.
\item if both $V(p_1),~V(p_2$) are non-zero, and $V(p_2)<V(p_1)$,
set $V(p_1)=V(p_2)$.
\item if both $V(p_1),~V(p_2$) are non-zero, and $V(p_1)=V(p_2)$,
do nothing.
\end{enumerate}
At the end of looping through the surface-pair links, set $N_v=n_v$. 
\item{VI.}  Repeat step V, except for the initializations and the
setting of $N_v$.  Continue iterating through the surface-pair links
until convergence is reached; i.e. there is no further modification of 
the $\{V(n)\}$. 
\item{VII.}  Initialize vortex areas $a_n=0,~n=1,...,N_v$.  Loop through
the negative plaquette number $n=1,...,A$.  At each plaquette, increment
\beq
a_{V(n)} = a_{V(n)} + 1
\eeq
\item{VIII.}  Eliminate any zero entries in the set of $a_n$.  This can
be done by setting $m=0$ and looping through index $n=1,...,N_v$.
If $a_n \ne 0$, increment $m=m+1$ and set $b_m=a_n$.  At the end
of the loop, reset $N_v=m$.  This is the total number of vortices. 
\item{IX.}  Calculate $s_w$.
\beq
       s_w = \sum_{m=1}^{N_v} \left( {b_m\over A} \right)^2
\eeq
\end{description}

\end{document}